%%%%%%%%%%%%%%%%%%%%%%%%%%%%%%%%%%%%%%%%%%%%%%%%%%%%%%%%%%%%%%%%%%%%%%%%%%%%
%%%%%%%%%%                                                  %%%%%%%%%%%%%%%%
%%%%%%%%%%             This paper is in PLAINTEX            %%%%%%%%%%%%%%%%
%%%%%%%%%%                                                  %%%%%%%%%%%%%%%%
%%%%%%%%%%%%%%%%%%%%%%%%%%%%%%%%%%%%%%%%%%%%%%%%%%%%%%%%%%%%%%%%%%%%%%%%%%%%

%: Page Setup for 8.5 x 11 inch paper 

 \message{Assuming 8.5 x 11 inch paper.}
 \message{' ' ' ' ' ' ' '}

\magnification=\magstep1	% \magstep1=1200
\raggedbottom

\parskip=9pt

\def\singlespace{\baselineskip=12pt}      % spacing for stuff like abstract
\def\sesquispace{\baselineskip=16pt}      % spacing for main text

%: Load or define some TeX macros 

%: This is the file `mathmacros.tex' 
%
%  Previously this was called `genmacros.tex' or `texmacros.tex'.
%
%  The macros here are mainly for mathematical symbols, while
%  the file `msmacros.tex' has most of the formatting macros for papers.
%
%                        Time-stamp:<2000 Jul 04  20:21:49 (14690 32669)> 

%:------------------------------------------------------------
%: For Spanish Accents do `tengeneza-tex-acentos'
%:-------------------------------------------------------------

%: Symbols in "openface" font for the integers, reals, etc.

%% The \hbox seems to be needed if the symbol is used in math mode.

 at10pt

\def\=>{\Rightarrow}
\def\==>{\Longrightarrow}

%: eric's d'Alembertian (good darkness, but a bit too small)
%  (but don't call it box since that exists already) 
 \def\dal{\displaystyle{{\hbox to 0pt{$\sqcup$\hss}}\sqcap}}

%-----------------------------------------------------------------------------

%: Symbols for "less than or of the order of" and its inverse 
%  taken from David Wiltshire

\def\lto{\mathop
        {\hbox{${\lower3.8pt\hbox{$<$}}\atop{\raise0.2pt\hbox{$\sim$}}$}}}
\def\gto{\mathop
        {\hbox{${\lower3.8pt\hbox{$>$}}\atop{\raise0.2pt\hbox{$\sim$}}$}}}
%
% Alternate versions
%
%   for a still better general method see arvinds  "stacksymbol" in the
%   file { ~/ms/texfiles/developing.macros/arvind.macros }
%
% \def\lto { {\raise1pt\hbox{$<$}} \!\!\!\! {\lower4pt\hbox{$\sim$}} }
% \def\gto { {\raise1pt\hbox{$>$}} \!\!\!\! {\lower4pt\hbox{$\sim$}} }
% \def\lto {{\lower4pt\hbox{$\buildrel<\over\sim$}}}
% \def\gto {{\lower4pt\hbox{$\buildrel>\over\sim$}}}

%-----------------------------------------------------------------------------

%: the fraction 1/2

%: More symbols

			% These seem more useful than the two
			% commented out ones just below
				% Could, also just redefine \langle and
				% \rangle to be these
	% \def\ket#1{|#1>}      %
	% \def\bra#1{<#1|}      %

		% symbol for isomorphism

		% symbol for set-theoretic difference

	% symbol used eg in f:A-->B

  %%%\def\to{\rightarrow}	

		% define tilde to always be the ``widetilde'' 
		% define bar to always be wide bar
		% define hat to always be the ``widehat'' 

		% triple equal sign

		% tensor product symbol

\def\interior #1 {  \buildrel\circ\over  #1}     % seems to work
 % alternate
 % \def\interior #1 {{ \buildrel\circ\over{{#1}} }} % works, not too well

% Lie Derivative symbol
  % notice that ${\rm{\it\$}}$ % fails

% semidirect product

%% These are for basis vectors and covectors, with labels (between
%% parentheses) directly over or under the ``kernel''.
%% The order of arguments is: kernel - label - vector index
%% usage example: \dualbasisvector{e}{j}{\mu}

\def\basisvector#1#2#3{
 \lower6pt\hbox{
  ${\buildrel{\displaystyle #1}\over{\scriptscriptstyle(#2)}}$}^#3}

%: END of the file `mathmacros'

% This is the file `msmacros.tex'
% Macros specifically for manuscripts, concerning such things as
% title, abstract, section titles, page formatting, etc.

                          % Time-stamp:<2000 Oct 04  17:19:16 (14811 40660)>

% (Outline "\\\\..." "%%..")

%% Following is to improve appearance of footnotes
%
 \let\miguu=\footnote
 \def\footnote#1#2{{$\,$\parindent=9pt\baselineskip=13pt%
 \miguu{#1}{#2\vskip -7truept}}}
%
% Notes
%   NOT breaking the line in the middle of the footnote-macro seems crucial!!??
%   Without the 9pt indent, the footnote symbol sticks out to the left.
%   The vskip at the end adjusts separation between multiple footnotes
%      (even 0pt separates them too much!)
% NB
%   Insert an extra \vskip -7pt (say) if phantom last line of footnote
%   spills over onto next page!
%
%-----------------------------------------------------------------------------

\def\BulletItem #1 {\item{$\bullet$}{#1}}

	% prevent page break (usually works)

\def\AbstractBegins
{
 \singlespace                                        % spacing for abstract
 \bigskip\leftskip=1.5truecm\rightskip=1.5truecm     % begin indentation
 \centerline{\bf Abstract}
 \smallskip
 \noindent	% this doesn't seem to take effect over a blank line
 } 
\def\AbstractEnds
{
 \bigskip\leftskip=0truecm\rightskip=0truecm
 }

\def\section #1 {\bigskip\noindent{\bf #1 }\par\nobreak\smallskip\noindent}

\def\subsection #1 {\medskip\noindent{\it [ #1 ]}\par\nobreak\smallskip}

\def\eprint #1 {{$\langle$e-print archive: #1$\rangle$}}

\def\linebreak{\hfil\break}

%% END of the file `msmacros'

%: preprint number(s)

\phantom{}
\vskip -1 true in
\medskip
\rightline{physics/0011002}
\vskip 0.3 true in

\vfill

\bigskip
\bigskip

%: Title 

\sesquispace
\centerline{ {\bf A HISTORICAL PERSPECTIVE ON CANCER}\footnote{*}%
{The main idea exposed herein occurred to me some 20
years ago.  I tried once to get it published, but failed.  
After some hesitation,
I am posting
it here in the hope that the historical viewpoint it utilizes may still
be of some use to people trying to understand cancer.}}

\bigskip

%: Authors 

\singlespace			        % (spacing for addresses etc.)

%:: FIRST AUTHOR 

\centerline {\it Rafael D. Sorkin}
\medskip
\smallskip
 
%:: FIRST ADDRESS

\centerline {\it Department of Physics, 
                 Syracuse University, 
                 Syracuse, NY 13244-1130, U.S.A.}

\smallskip
 
%:: EMAIL ADDRESS

\centerline {\it \qquad\qquad internet address: sorkin@physics.syr.edu}

\AbstractBegins 
It is proposed that cancer results from the breakdown of universal
control mechanisms which developed in mutual association as part of the
historical process that brought individual cells together into
multi-cellular communities.  By systematically comparing the genomes of
uni-celled with multi-celled organisms, one might be able to identify
the most promising sites for intervention aimed at restoring the damaged
control mechanisms and thereby arresting the cancer.
\AbstractEnds

%: Set spacing for body of paper

\sesquispace

\bigskip\medskip

%: First block of main text 

More or less by definition, a cancerous cell is one that grows and
reproduces uncontrollably.  Of course, this characterization presupposes that
the cell in question lives within a multi-celled organism --- a
community of cells.  The identical behavior would appear perfectly
normal for a cell existing in isolation.  

But there is more to cancer than uncontrolled growth; for this behavior
occurs in association with another trait which --- in itself --- would not
seem to be related to growth at all: loss of differentiation.  Why
should a ``histologic'' characteristic like (lack of) differentiation be
correlated in this way with a ``dynamical'' characteristic like growth
rate?
I believe that, far from being necessary for reasons of biochemistry or
cellular dynamics, this association is {\it historical} in origin. 

%% I believe that there lies within this question a cruccial hinbt to to
%% the origins of -- and possibly a cure for -- this disease

Let us try to imagine the transformation that took place when the first
multicellular organisms coalesced from collections of cells living in
relative isolation from each other.  Aside from the mutual
``solidarity'' of its members, such a community of cells has the
advantage that different cells can specialize in different tasks and
collectively 
accomplish all these tasks more
effectively than any individual cell could by itself.  In other words, a
division of labor becomes possible (specialized tissues, specialized organs,
etc.)
But for such an arrangement to function, 
the relative multiplicities
of the cells with different specializations must be suitable, which
in practice means controlling their absolute numbers as well.  To
participate in such a community, then, a cell needs ({\it at least}) 
the ability to specialize itself for a range of functions 
and 
the ability to regulate its rate of reproduction as needed: 
precisely the features whose absence typifies the disease of cancer.

It seems clear that, 
in order to provide these twin abilities, new biological
``machinery'' would have been required, either at the cellular level, or
the supercellular level, or both.  Furthermore, since the required
mechanisms must have arisen in association with 
one another 
as part of
the same historical process 
(a process presumably occupying an extended period of time), 
it would be natural for them to overlap and share components to a
great extent.  That is, there should have developed,
in some measure, 
only a single
(and universal) mechanism responsible on one hand for regulating growth
and on the other hand for producing functional differentiation.
Cancer would then result from the breakdown of this mechanism.
From this perspective, 
cancer would not be 
a disease 
like measles,
caused by the {\it presence} of some pathogen 
or some actively harmful abnormality, 
but  
a disease of deficiency 
like scurvy,
caused by the {\it absence} or the failure 
of some mechanism that is present in normal cells (or normal tissue).  
It would be the consequent 
{\it regression}, 
on the part of the tumor cells,
to an earlier, pre-social form of behavior.
 
% Certain testable implications follow from this historical hypothesis and
% certain therapeutic consequences as well.  The latter constitute, not any
% definite prescription for a cure, but rather a particular
% outlook on the problem.

The focus for finding a cure 
(or for prevention)
would thus shift from {\it removing} the
abnormality to {\it restoring} the missing or damaged machinery, 
and
a strategy of trying to destroy the defective cells would be 
less likely to succeed 
than 
one of providing them with ``replacement parts''
for their damaged control mechanisms.
How this could be done would naturally depend on the nature of the
defect, but one might imagine, for example,
delivering the replacement parts
by means of retro-viruses or similar agents.
With luck, such replacement
parts
would not harm healthy tissue, and hence 
they
(unlike traditional chemotherapeutic agents) 
could be introduced 
in numbers sufficient to reach all the cells they needed to reach.

Before it could be repaired, however, 
the relevant ``machinery'' would have to be identified.  
Here a plausible concomitant of the postulated historical transition to
communal life becomes relevant, 
namely 
the {\it universality} of the control mechanisms that evolved then.
Assuming that the transition to multi-celled organisms was a prolonged
process, one would expect the control mechanisms that developed in its course
to be shared by all, or almost all, multi-celled forms.  Conversely,
uni-celled organisms should lack these control mechanisms.  

To test this idea,
one
could compare the genomes of one-celled organisms with those of 
multi-celled organisms susceptible to cancer, 
searching for genes, 
or sequences thereof, that were universally present in the latter
and absent in the former.  
These genes (or sequences)
would then be
the ones responsible for the postulated control mechanisms.  
Moreover, the {\it oldest} such genes 
(or sequences), would furnish better candidates than the newer ones,
assuming that age could be identified reliably.
Any attempt to develop a ``gene therapy'' or diagnostic test for
cancer could then focus on the components of the genome
identified in this way.

Of course, it might be overly optimistic to expect that the postulated
control mechanisms could be identified simply with some well delineated
portion of a cell's genome.  They might involve non-genetic components
whose failure was not occasioned by genetic mutations.  
On the other hand, 
one knows that 
genetic change is involved in cancer and that, 
in particular, 
there is a close correlation between mutagenicity and
carcinogenicity.  Given these facts, 
it does not seem too much to hope that 
the required ``social skills'' are coded into
universal portions of the genome
which could be identified and catalogued. 
Conversely, the failure to discover
such portions would constitute evidence {\it against} the historical
explanation of cancer put forward in this article.

%: Acknowledgments including grant citation

\bigskip\noindent

This research was partly supported by NSF grant PHY-9600620 
and by 
a grant from the Office of Research and Computing of Syracuse University.

\end